\documentclass{aa}  

\usepackage{graphicx}
\usepackage{txfonts}
\usepackage{lipsum}
\usepackage{subcaption}         % necessary for continued figures, example in section 3
\usepackage{lscape}             % to rotate a single page table, example in appendix.
\usepackage{placeins}           % useful with \FloatBarrier, to keep 
\usepackage{siunitx}    % A comprehensive (si) units package

\usepackage[version=4]{mhchem}
\usepackage{xspace}

\usepackage{booktabs}

\usepackage{makecell}
\usepackage{multirow}
\usepackage{multicol}
\usepackage{pdflscape}	% Landscape pages

\begin{document}

   \title{Soft X-ray line emission from hot gas in intervening galaxy halos and diffuse gas in the cosmic web}

   \author{Yuning Zhang\inst{1}\fnmsep\thanks{E-mail: zhangyuning@mail.tsinghua.edu.cn}
        \and Dandan Xu\inst{1}\fnmsep\thanks{E-mail: dandanxu@tsinghua.edu.cn}
        \and Chengzhe Li\inst{1}
        \and Wei Cui\inst{1}
        }

   \institute{Department of Astronomy, Tsinghua University, Beijing 100084, China}

   \date{Received September 30, 20XX}

  \abstract
   {Cosmic hot-gas emission is closely related to halo gas acquisition and galactic feedback processes. Their X-ray observations reveal important physical properties and movements of the baryonic cycle of galactic ecosystems. However, the measured emissions toward a target at a cosmological distance would always include contributions from hot gases along the entire line of sight to the target. Observationally, such contaminations are routinely subtracted via different strategies. With this work, we aim to answer an interesting theoretical question regarding the amount of soft X-ray line emissions from intervening hot gases of different origins. We tackled this problem with the aid of the TNG100 simulation. We generated typical wide-field light cones and estimated their impacts on spectral and flux measurements toward X-ray-emitting galaxy-, group- and cluster-halo targets at lower redshifts. We split the intervening hot gases into three categories; that is, the hot gas that is gravitationally bound to either star-forming or quenched galaxy halos, and the diffuse gas, which is more tenuously distributed permeating the cosmic web structures. We find that along a given line of sight, the diffuse gas that permeates the cosmic web structures produces strong oxygen and iron line emissions at different redshifts. The diffuse gas emission in the soft X-ray band can be equal to the emission from hot gases that are gravitationally bound to intervening galaxy halos. The hot-gas emission from the quiescent galaxy halos can be significantly less than that from star-forming halos along the line of sight. The fluxes from all of the line-of-sight emitters as measured in the energy band of $\SIrange{0.4}{0.85}{keV}$ can reach $\sim \SIrange{20}{200}{\percent}$ of the emission from the target galaxy, group, and cluster halos. The fluxes from the intervening hot gas as measured in narrow bands around the \ion{\ce{O}}{\romannumeral7} (r) and \ion{\ce{O}}{\romannumeral8} (K$\alpha$) are typically only a few percent of the target emission, indicating that these line emissions (as measured within narrow bands) better present the hot-gas emission of the target sources, compared to that measured in wider energy bands. }

   \keywords{methods: numerical --
            galaxies: halos --
            galaxies: intergalactic medium --
            X-rays: galaxies --
            techniques: spectroscopic
            }

   \maketitle
   
   \nolinenumbers

%%%%%%%%%%%%%%%%%%%%%%%%%%%%%%%%%%%%%%%%%%%%%%%%%%%%%%%%%%%%%%
\section{Introduction}

Hot gas inside gravitationally bound dark-matter halos and extensively permeating the cosmic web structures, is mainly produced by gravitational heating during cosmic structure formation \citep[e.g.,][]{2005MNRAS.363....2K, 2006MNRAS.368....2D, 2009Natur.457..451D}, or by baryonic feedback processes during galaxy evolution; the typical temperature range is $\SI{e5}{K} \lesssim T \lesssim \SI{e7}{K}$. In particular, the heating process is directly linked to various feedback mechanisms of baryonic physics, which eject mass, energy, and momentum to large distances from the galaxy centers, causing galaxy halos suffering from gas depletion at different levels \citep[e.g.,][]{2000MNRAS.317..697E, 2006ApJ...650..560C, 2007MNRAS.374..427D, 2014ARA&A..52..529Y, 2018ApJ...857..121Y, 2016MNRAS.457.1385W, 2021MNRAS.501.3640H, 2023ARA&A..61..131F, 2025MNRAS.536.3200P}. Theories and simulations have revealed that half of the cosmic baryons are in fact in the form of diffuse hot gas and reside in the large-scale structures. However, this diffuse hot gas component is extremely challenging to observe (\citealt{1998ApJ...503..518F, 1999ApJ...514....1C, 2001ApJ...552..473D, 2005ApJ...620...21K, 2006ApJ...650..573C, 2007ARA&A..45..221B, 2010MNRAS.408.2051D, 2012ApJ...759...23S, 2016MNRAS.457.3024H, 2019MNRAS.486.3766M}).

In the past decades, cosmic and galactic hot gas has become a topic of growing importance in the field of structure formation and evolution. This is a result of combined advances in both observations and theories through the help of simulations. Current-generation X-ray missions, such as \textit{Chandra}, \textit{XMM-Newton}, and \textit{eROSITA} \citep{2021A&A...647A...1P}, already revealed rich and detailed properties and motions of hot gas in massive galaxies, groups and clusters \citep{2020A&A...633A..42S, 2014Natur.515...85Z, 2019NatAs...3..832Z}, as well as between galaxy clusters (e.g., gas bridges detected through the Sunyaev-Zel'dovich effect in combination with X-ray observations \citep{2013A&A...550A.134P} or direct imaging from X-ray observations \citep{2020MNRAS.491.2605P}). Next-generation X-ray missions, such as the X-Ray Imaging and Spectroscopy Mission (\textit{XRISM}; \citealt{2020arXiv200304962X}), the Advanced Telescope for High-ENergy Astrophysics (\textit{Athena}; \citealt{2013arXiv1306.2307N}), the Diffuse Intergalactic Oxygen Surveyor (\textit{Super DIOS}; \citealt{2018JLTP..193.1016Y}), and the Hot Universe Baryon Surveyor (\textit{HUBS}; \citealt{2020JLTP..199..502C, 2020SPIE11444E..2SC}), have either been put in place or are in the planning stage. Equipped with microcalorimeter-based detector arrays and large mirrors, these next-generation X-ray telescopes with even higher energy resolution ($\si{eV}$-level) and detection efficiencies are dedicated to pushing current detection limits to a further extent. 

Our theoretical understanding of hot-gas physics has also been making steady progress, in particular since the advent of state-of-the-art cosmological hydrodynamical simulations \citep[e.g.,][]{2010MNRAS.407.1403C, 2013MNRAS.430.2688V, 2018MNRAS.477..450N, 2020MNRAS.491.2939O, 2020ApJ...893L..24O, 2022MNRAS.514.5214W, 2023MNRAS.522.3665N}, including those recent dedicated cluster simulation suites within a cosmological context (TNG-Cluster simulation; see \citealt{2024A&A...686A.157N, 2024A&A...687A.129L, 2024A&A...686A.200T, 2024A&A...686A..86R, 2024A&A...690A..20A}). 

\citet{2020ApJ...893L..24O} made specific mock observations for the \textit{eROSITA} instrument on the Spectrum-Roentgen-Gamma mission (\citealt{2012arXiv1209.3114M}) using the IllustrisTNG (\citealt{2018MNRAS.475..648P, 2018MNRAS.480.5113M, 2018MNRAS.475..624N, 2018MNRAS.475..676S, 2018MNRAS.477.1206N}) and the EAGLE (\citealt{2015MNRAS.446..521S}) simulations. They found that the hot circumgalactic-medium (CGM) emission in central galaxies with lower masses and a higher specific star formation rate (sSFR) can be detected out to a distance of $\SIrange{30}{50}{kpc}$, while in higher mass central galaxies this detection can go up to $\SIrange{150}{200}{kpc}$. Both simulations also predicted higher soft X-ray luminosities (at fixed stellar mass) for higher sSFR galaxies, in comparison to their lower sSFR galaxy counterparts. Using three suits of hydrodynamical simulations, that is, EAGLE, IllustrisTNG, and SIMBA (\citealt{2019MNRAS.486.2827D}), \citet{2024ApJ...969...85S} demonstrated the feasibility of characterizing the CGM in nearby galaxy halos ($z\sim \numrange{0.01}{0.03}$) using next-generation X-ray microcalorimeter. Through the mock spectroscopic observations made with typical instrumental design parameters, they found that individual halos of Milky Way mass can be traced out to large distances using prominent emission lines, such as \ion{\ce{O}}{\romannumeral7}, \ion{\ce{O}}{\romannumeral8} and iron lines. They also demonstrate the ability of X-ray observations to reveal the spatial distributions of temperature, velocity, and abundance ratio through spectral fitting for individual galaxy halos. 

We note that an observationally accurate and clean measurement of any given target X-ray source at a cosmological distance is not an easy task, simply because the total detected emission also includes contamination from the Milky Way foreground \citep[e.g.,][]{2023A&A...670A..99P, 2024A&A...681A..78L} as well as from all emitters along the entire line of sight. In light of next-generation X-ray missions equipped with microcalorimeters, our previous study (\citealt{2022ExA....53.1053Z}) showed that with a spectral resolution as high as $\SI{2}{eV}$, the bright X-ray emission lines (such as \ion{\ce{O}}{\romannumeral7} and \ion{\ce{O}}{\romannumeral8}) from targets at redshifts beyond $z\sim 0.01$ will be able to be separated from the Milky Way foreground emissions. This is also highlighted by \citet{2024ApJ...969...85S}, which showed the necessity of implementing microcalorimeters at a spectral resolution as high as $\SI{2}{eV}$ to distinguish the faint CGM line emission from the target sources from the bright Milky Way foreground emission. 

Regarding the line-of-sight contamination, \citet{2022ApJ...936L..15C} detected spatially resolved CGMs in both star-forming and quiescent galaxies by stacking the X-ray observations of optically selected galaxies between $0.01 < z < 0.1$ in the \textit{eROSITA} Final Equatorial Depth Survey (eFEDS), and the results were compared to the IllustrisTNG and Eagle cosmological simulations. In order to subtract contribution from the hot halo emission originated in groups and clusters that lie along the line of sight either in the foreground or background of the targeted galaxies, they removed these extended sources by cross-matching the eFEDS group and cluster catalog (\citealt{2022A&A...661A...2L}), which contains $542$ candidate groups and clusters in the redshift range of $0.01$ to $1.3$, with a median redshift of $0.35$. It turned out that only $37$ of $2643$ targeted galaxies lie within a projected distance of twice $R_{500}$ from a group or cluster. To further remove X-ray emission from diffuse hot gas along the line of sight, they subtracted from the stacked signal the median of the average emission within an annulus of $\SIrange{150}{300}{kpc}$ from each galaxy in the stacked sample, assuming that the target emission does not go beyond $\SI{100}{kpc}$. Using a similar approach through stacking the X-ray emission from galaxies in the \textit{SRG/eROSITA} All-Sky Surveys, \citet{2024A&A...690A.267Z} detected hot CGM in Milky Way-mass and M31-mass galaxy halos out to $z=0.2$ and found that the baryon budget in the hot halo gas is lower than the $\Lambda$ Cold Dark Matter ($\Lambda$CDM) prediction, indicating significant gas depletion in halos at these scales. In their study, contamination effects from satellite galaxies were dealt with by carefully constructing galaxy samples using spectroscopic redshifts and comparing to cosmology simulations. 

The primary goal of this study is, using cosmological simulations and identifying and characterizing the hot gas across a range of redshifts, to theoretically explore the significance of soft X-ray emission from different intervening components along some typical line of sight in light of high-energy resolution soft X-ray spectroscopic observations. In particular, we compared the line-of-sight emissions from the diffuse gas components, star-forming galaxy halos, and quiescent galaxy halos, and we also compared the total emissions from the entire line-of-sight emitters to the emission from the target galaxy, group, and cluster halos. We find that the diffuse gas produces strong oxygen and iron line emissions at different redshifts, contributing almost equally to hot gas that is gravitationally bound to intervening galaxy halos; while star-forming galaxy halos contribute more line emissions than quiescent galaxy halos along the line of sight. In general, line emissions as measured in narrow bands around \ion{\ce{O}}{\romannumeral7} (r) and \ion{\ce{O}}{\romannumeral8} (K$\alpha$) can better present the hot-gas emission of the target sources, as the contributions from the line-of-sight emitters typically represent a few percent of the target emission; on the other hand, the line-of-sight emission measured in broader soft X-ray energy bands can reach several tens of percent of the target emission. 

The paper is structured as follows. In Sect.~\ref{sec:methods}, we present the simulation data, the method of producing the light cone, the classification of the gas, and how to calculate the X-ray emission. The results on the gas in the light cone are shown in Sect.~\ref{sec:results}. Finally, the discussion and the conclusions of this work are given in Sect.~\ref{sec:DiscussionConclusions}. 

%%%%%%%%%%%%%%%%%%%%%%%%%%%%%%%%%%%%%%%%%%%%%%%%%%%%%%%%%%%%%%
\section{Methods}
\label{sec:methods}

\subsection{Numerical simulation}

In this work, we used data from the IllustrisTNG project\footnote{\url{http://www.tng-project.org/}} (TNG hereafter; \citealt{2018MNRAS.475..648P, 2018MNRAS.480.5113M, 2018MNRAS.475..624N, 2018MNRAS.475..676S, 2018MNRAS.477.1206N}), which is one of the state-of-the-art, large cosmological magnetohydrodynamical simulations. The main scientific goal of the TNG project is to shed light on the key physical processes of cosmic structure formation and evolution. The TNG simulation was performed with a parallel moving-mesh code \textsc{arepo} \citep{2010MNRAS.401..791S}. The galaxy formation model of TNG includes gas radiative cooling, star formation, stellar evolution, chemical enrichment, black-hole formation, growth, and multi-mode feedback \citep{2017MNRAS.465.3291W, 2018MNRAS.473.4077P}. The TNG simulation suite includes three primary runs: TNG50, TNG100, and TNG300. Each run was carried out at three resolution levels. In this work, we adopted the full baryonic physics data of TNG100-1, which are moderate in size (the box side length is $L_{\mathrm{box}} = \SI{110.7}{Mpc}$) and high resolution (the dark-matter and mean baryon particle-mass resolutions are $m_{\mathrm{DM}} = \SI{7.5e6}{M_{\sun}}$ and $m_{\mathrm{baryon}} = \SI{1.4e6}{M_{\sun}}$, respectively). Dark-mater halos and galaxies hosted within were identified using the friends-of-friends (\textsc{FoF}) \citep{1985ApJ...292..371D} and \textsc{subfind} \citep{2001MNRAS.328..726S, 2009MNRAS.399..497D} algorithms, respectively. The simulation and this work adopt the standard $\Lambda$CDM cosmological parameters advanced by the \citet{2016A&A...594A..13P}, namely a dark-energy density parameter, $\Omega_{\Lambda} = 0.6911$; matter density parameter, $\Omega_{\mathrm{m}} = 0.3089$; baryon density parameter, $\Omega_{\mathrm{b}} = 0.0486$; power spectrum normalization characterized by $\sigma_8 = 0.8159$; scalar spectral index, $n_\mathrm{s} = 0.9667$; and dimensionless Hubble constant, $h = 0.6774$. 

The TNG data releases include the snapshots and halo (subhalo) catalogs at each snapshot \citep{2019ComAC...6....2N}. The halo (subhalo) catalogs include \textsc{FoF} and \textsc{subfind} objects. The snapshot data include all simulation particles and cells. We note that any given \textsc{FoF} halo may contain one or more subhalos. The particles (cells) belonging to subhalos are gravitationally bound. While the diffuse and gravitationally unbound particles (cells) are in general referred to as ``fuzz'' elements. Depending on whether they belong to any \textsc{FoF} halo, the ``fuzz'' particles (cells) can be further divided into the ``inner fuzz'' or the ``outer fuzz'' categories. The gas cells belonging to these different categories were used to construct our light cones. We note that this ``fuzz'' component physically includes all hot gas that is not gravitationally bound to any galaxies and dark-matter halos, including the intracluster medium (ICM), intragroup medium (IGrM), and intergalactic medium (IGM) permeating the cosmic web. The ICM is the diffuse hot gas that resides in a cluster of galaxies but does not belong to any substructures of the cluster \citep{ryden_pogge_2021}. IGrM is similar to ICM but for group of galaxies. In simulation, these types of gases can be classified as inner-fuzz cells that belong to a \textsc{FoF} group but are not bound to any \textsc{subfind} subhalo. On the contrary, the IGM can be well approximated by the outer fuzz in the simulation, since this diffuse gas component does not belong to any groups/halos. 

\subsection{Simulated wide-field light cone}

In order to study the contribution of X-ray emission from hot gas along any given line of sight but outside a target source, we constructed a simulated light cone. We can link the discrete redshift outputs of the numerical simulation with practical observations. In particular, to study the properties of the hot gas in the filamentary structures, a wide-field light cone is needed. 

In this work, we adopted the method of \citet{2007MNRAS.376....2K} and \citet{2022ExA....53.1053Z} to construct a simulated light cone without repeated structures. The key to avoiding the kaleidoscopic effect due to repeated usage and periodicity of the simulation box is to choose a pair of integers $(n,m)$ that do not have common denominator. Once a pair of integers $(n,m)$ without a common factor were chosen, the corresponding line of sight, field of view, and observation redshift range were uniquely determined.
Considering the box size of TNG100, we chose a pair of integers $(n,m) = (4,3)$ in order to obtain the simulated wide-field light cone with a FoV of $\sim1.59\degr \times 1.19\degr$ and redshift range of $z = \numrange{0}{0.356}$. We note that only the central square area with the field of $1\degr \times 1\degr$ was treated as the angular coverage of the light cone and used for our statistics. In addition, this redshift range is also far beyond the target galaxy, group, and cluster halos at cosmological distances that are expected and plausible to observe using the next-generation X-ray telescopes, such as \textit{HUBS} (see \citealt{2022ExA....53.1053Z}). To construct the hot-gas emission in a given light cone, we took snapshots at $z=0.01$, $0.03$, $0.06$, $0.11$, $0.18$, $0.27$, and $0.35$. This allowed us to take into account the evolution of the large-scale structure in a self-consistent way. Further through a serious of normal transformations (i.e., swapping the integers, changing the origin, and permuting coordinate axes), we generated 48 light cones under the chosen integer pair set. In this study, we took nine which contain no overwhelmingly bright structures to dominate the light-cone emission and used them for a general number-counting statistical investigation (see Table~\ref{tab:gal}). Due to computational cost, we adopted one (LC0) light cone for a detailed calculation and evaluation of the hot-gas emission properties (see Sect.~\ref{sec:results}).

\subsection{Component identification}

Hot gas within any given line of sight but outside the target source has two major origins: gravitationally bound galaxies and their dark-matter halos, and diffuse gas in the intergalactic medium between galaxy halos. Apart from evaluating their X-ray emission in the light cone and making comparisons to emission from the target source, it is also interesting to compare the X-ray emission from hot gas of different origins. For this reason, we categorized the gas cells in the light cone into several types corresponding to different observational characteristics. 

\begin{figure}
  \centering
  \includegraphics[width=\columnwidth]{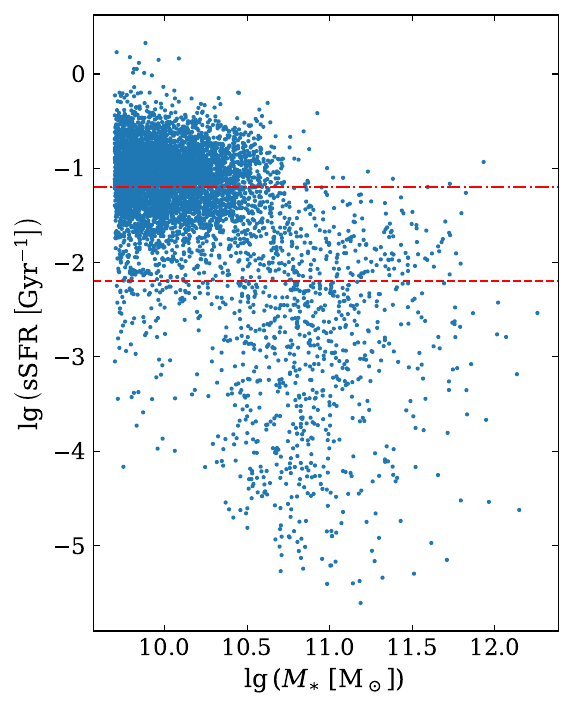}
  \caption{Specific star formation rate (sSFR) as a function of stellar mass for all galaxies at $z=0.01$ (snapshot 98). The stellar mass here was calculated using all stellar particles within twice the stellar half-mass radius of a given subhalo. The red dash-dotted line represents the mean sSFR. The red dashed line represents $\SI{1}{dex}$ below the mean sSFR, below which galaxies are classified as quenched galaxy populations \citep[e.g.,][]{2021MNRAS.503..726L}. }
  \label{fig:sSFR}
\end{figure}

First, we considered the hot gas that is gravitationally bound to the galaxy halos. To guarantee that the galaxies have enough baryonic mass to be sufficiently resolved, we only selected the subhalos with a stellar mass of $M_{\ast} > \SI{5e9}{M_{\sun}}$, where $M_{\ast}$ is the sum of masses of all stars within twice the stellar half-mass radius. We note that only $\SI{5.33}{\percent}$ of the total gas mass resides in subhalos below this stellar-mass threshold. According to their star formation activities, we further split the selected galaxies into a subsample of star-forming galaxies and a subsample of quenched galaxies. The method we used to identify different types of galaxies is similar to the practices of \citet{2018MNRAS.474.3976G} and \citet{2021MNRAS.503..726L}, which we roughly summarize below. We calculated the mean specific star formation rate ($\mathrm{sSFR} = \mathrm{SFR}/M_{\ast}$, where $\mathrm{SFR}$ is the star formation rate) within twice the stellar half-mass radius of these subhalos in a snapshot. The sSFR of a subhalo as a function of the stellar mass at $z=0.01$ is shown in Fig.~\ref{fig:sSFR}. Galaxies that lie below $\SI{1}{dex}$ below the mean sSFR are classified as quenched galaxies. We refer to galaxies above this threshold as star-forming galaxies. Such a classification was consistently carried out at different redshifts for this study. In addition, we also identified a diffuse gas component in the light cone as the collection of all ``fuzz'' gas cells, that is, including the ICM, IGrM (classified as ``inner fuzz''), and IGM (``outer fuzz'' in the simulation). 

Table~\ref{tab:gal} lists the total numbers of star-forming and quenched galaxies, and the total amount of stellar mass and gas mass, contained in the investigated light cone (LC0) of a $1\degr \times 1\degr$ field of view. It can be seen that the number of star-forming galaxies is significantly greater than that of quenched galaxies. However, the total stellar mass and total gas mass are similar among the two types of galaxies, as the latter samples are more massive than the former in general. In comparison, the diffuse gas component takes up the highest mass fraction in comparison to gas bound to galaxies; i.e., about five times the latter. This finding is broadly consistent with the study of \citet{2021A&A...649A.117G}, which found that the warm-hot intergalactic medium in filament accounts for $\sim \SI{80}{\percent}$ of the local baryon budget using the TNG300 simulation. With such a component being the major part of the diffuse gas, it is also the primary candidate for the ``missing baryons'' in the local Universe \citep[e.g.,][]{1999ApJ...514....1C, 2001ApJ...554L...9P, 2003PASJ...55..879Y, 2005Natur.433..495N, 2008Sci...319...55N, 2013arXiv1306.2324K, 2018Natur.558..406N}. 

We note that, given the investigated field of view, the hot-gas emission can markedly differ among different lines of sight. In principle, one can generate a sample of light cones for such analyses in order to better assess the scatter. However, as the calculation of the hot-gas emission is already extensive for one light cone, we also generated another eight different light cones only to examine the numbers of intervening galaxies, and the amount of baryonic masses in these light cones for comparisons. We did not carry out any further calculations for the hot-gas emission therein. The statistical results are also listed in Table~\ref{tab:gal}. As can be seen, LC0 is a line of sight that is on the lower side in terms of the number of intervening galaxies and the amount of mass in baryons. In the next sections, we present the results of the hot-gas emission in the case of LC0. 

\begin{table*}
	\centering
	\caption{Summary of star-forming and quenched galaxies and diffuse gas per square degree in the simulated light cones (constructed up to $z=0.356$).}
	\label{tab:gal}
	\begin{tabular}{clS[table-format = 4]S[table-format = 1.3e2]S[table-format = 1.3e2]}
		\hline
             & \multirow{2}{*}{Type} & {\multirow{2}{*}{Number of galaxies}} & {Total stellar mass} & {Total gas mass} \\
             & & & {$\left(\si{M_{\sun}}\right)$} & {$\left(\si{M_{\sun}}\right)$} \\
		\hline
		\multirow{3}{*}{LC0} & Star-forming galaxies & 1174 & 3.299e13 & 1.114e14 \\
             & Quenched galaxies & 459 & 3.847e13 & 1.332e14 \\
             & Diffuse gas & \ldots & \ldots & 1.324e15 \\
		\hline
		\multirow{3}{*}{LC1} & Star-forming galaxies & 1217 & 3.426e13 & 1.283e14 \\
             & Quenched galaxies & 669 & 5.376e13 & 2.290e14 \\
             & Diffuse gas & \ldots & \ldots & 1.418e15 \\  
            \hline
		\multirow{3}{*}{LC2} & Star-forming galaxies & 1306 & 3.830e13 & 1.450e14 \\
             & Quenched galaxies & 675 & 5.395e13 & 2.448e14 \\
             & Diffuse gas & \ldots & \ldots & 1.410e15 \\  
            \hline
		\multirow{3}{*}{LC3} & Star-forming galaxies & 1683 & 5.236e13 & 2.004e14 \\
             & Quenched galaxies & 868 & 6.501e13 & 2.369e14 \\
             & Diffuse gas & \ldots & \ldots & 1.711e15 \\  
            \hline  
		\multirow{3}{*}{LC4} & Star-forming galaxies & 1272 & 3.852e13 & 1.641e14 \\
             & Quenched galaxies & 577 & 4.175e13 & 1.317e14 \\
             & Diffuse gas & \ldots & \ldots & 1.421e15 \\  
            \hline       
		\multirow{3}{*}{LC5} & Star-forming galaxies & 1222 & 3.539e13 & 1.418e14 \\
             & Quenched galaxies & 518 & 4.242e13 & 1.559e14 \\
             & Diffuse gas & \ldots & \ldots & 1.343e15 \\  
            \hline      
		\multirow{3}{*}{LC6} & Star-forming galaxies & 1300 & 3.412e13 & 1.241e14 \\
             & Quenched galaxies & 554 & 4.419e13 & 1.599e14 \\
             & Diffuse gas & \ldots & \ldots & 1.411e15 \\  
            \hline            
		\multirow{3}{*}{LC7} & Star-forming galaxies & 1389 & 4.320e13 & 1.684e14 \\
             & Quenched galaxies & 666 & 5.162e13 & 1.852e14 \\
             & Diffuse gas & \ldots & \ldots & 1.481e15 \\  
            \hline   
		\multirow{3}{*}{LC8} & Star-forming galaxies & 1427 & 4.227e13 & 1.657e14 \\
             & Quenched galaxies & 665 & 5.142e13 & 1.598e14 \\
             & Diffuse gas & \ldots & \ldots & 1.490e15 \\  
            \hline            
    \end{tabular}
    \flushleft
    Note: For the diffuse gas in the third line in each light-cone row, the last column (total gas mass) is calculated as the sum of masses of all the fuzz gas particles in the light cone. For either star-formation galaxies or quenched galaxies in the first or second line in each light-cone row, the total gas mass and the total stellar mass, respectively, are calculated as sums of all gas masses and stellar masses of subhalos located in a given light cone. Counting particles (belonging to subhalos) instead yields slightly lower values due to the boundary cut-off of the light cone. 
\end{table*}

\subsection{X-ray emission calculation}

The hot X-ray-emitting gas is usually assumed to be optically thin and in collisional ionization equilibrium (CIE). To model this gas, we adopted the Astrophysical Plasma Emission Code (\textsc{apec} v3.0.9; \citealt{2001ApJ...556L..91S, 2012ApJ...756..128F}). Using the Simulated Observations of X-ray Source (\textsc{SOXS}; \citealt{2023ascl.soft01024Z}) package, we calculated the X-ray emission of the gas with the \textsc{apec} model. For each gas cell, to derive its X-ray emission, we input its local density, temperature, and metallicity into the code. It should be noted that the metallicity value should be converted to solar metallicity by dividing by $0.0127$ (the primordial solar metallicity used in TNG; \citealt{2009MNRAS.399..574W}). Furthermore, the X-ray emission should be Doppler-shifted and cosmologically redshifted according to the peculiar velocity and the redshift of the gas cell. In practice, the soft X-ray energy band on which we focused is $\SIrange{0.1}{2}{keV}$. This band includes many interesting metal lines, such as, \ion{\ce{O}}{\romannumeral7} triplets ($\SI{0.561}{keV}$, $\SI{0.569}{keV}$ and $\SI{0.574}{keV}$), \ion{\ce{O}}{\romannumeral8} K$\alpha$ line ($\SI{0.654}{keV}$), and a series of \ion{\ce{Fe}}{\romannumeral17} lines (e.g., $\SI{0.725}{keV}$, $\SI{0.727}{keV}$, $\SI{0.739}{keV}$, and $\SI{0.826}{keV}$). 

% % %%%%%%%%%%%%%%%%%%%%%%%%%%%%%%%%%%%%%%%%%%%%%%%%%%%%%%%%%%%%%%
\section{Results}
\label{sec:results}

The hot gas in dark-matter halos is multiphase \citep{2012ARA&A..50..491P, 2017ARA&A..55..389T, 2020ARA&A..58..363P}. In our previous study (\citealt{2022ExA....53.1053Z}), we used the TNG100 simulation to investigate the soft X-ray-emission properties of hot gases in galaxy-, group- and cluster-sized dark-matter halos. We demonstrated that the hot gas of different phases and with different origins broadly spreads out in terms of its observational characteristics. In that study, we made exact mock observations of the hot CGM emission therein for the \textit{HUBS} mission\footnote{\url{https://hubs-mission.cn/en/index.html}} (\citealt{2020JLTP..199..502C, 2020SPIE11444E..2SC}). This process included generating X-ray spectra from these targets using the APEC code and sampling the spectra to make mock observations using PyXSIM, considering point-like emissions from an unresolved AGN population as the main component of the cosmic X-ray background, as well as adding the Milky Way foreground emission using the SOXS package. In this study, we do not aim to make any exact mock observations with all realistic components for any specific X-ray missions, but to evaluate (from a theoretical perspective) the amount of soft X-ray emission from hot gases in hundreds and thousands of galaxies as well as from the warm-hot intergalactic medium that permeates the cosmic web along typical lines of sight toward interesting targets. 

\subsection{Emissivity maps}

In order to evaluate the contribution of line-of-sight emissions and make comparisons to the target source emission, in this work, we used three types of X-ray targets that were previously studied in \citet{2022ExA....53.1053Z}. The three targets are a galaxy halo with a total mass of $M = \SI{3.06e12}{M_{\sun}}$ at $z=0.03$, a group halo of $M = \SI{7.00e13}{M_{\sun}}$ at $z=0.11$, and a cluster halo of $M = \SI{3.78e14}{M_{\sun}}$ at $z=0.11$. We note that the X-ray emission properties of dark halos depend on baryonic physical processes in particular the AGN feedback, and therefore they can vary among different halos, as well as among different simulations that treat baryonic physics processes differently (e.g., \citealt{2019MNRAS.485.3783D, 2023MNRAS.525.1976T, 2024A&A...690A.268Z}). The three dark halos selected from TNG100 are among typical soft X-ray emission sources with luminosities of $\SIrange{e40}{e43}{erg.s^{-1}}$ in $\SIrange{0.1}{2}{keV}$, and are ideal for soft X-ray imaging and spectroscopic studies (\citealt{2022ExA....53.1053Z}). As shown by \citet{2024A&A...690A.268Z}, the TNG100 simulation is roughly consistent with the \textit{eROSITA} observations (within $\SI{1}{dex}$) in terms of the scaling relations between X-ray luminosity and galaxy/halo mass.

\begin{figure*}
  \centering
  \includegraphics[width=\linewidth]{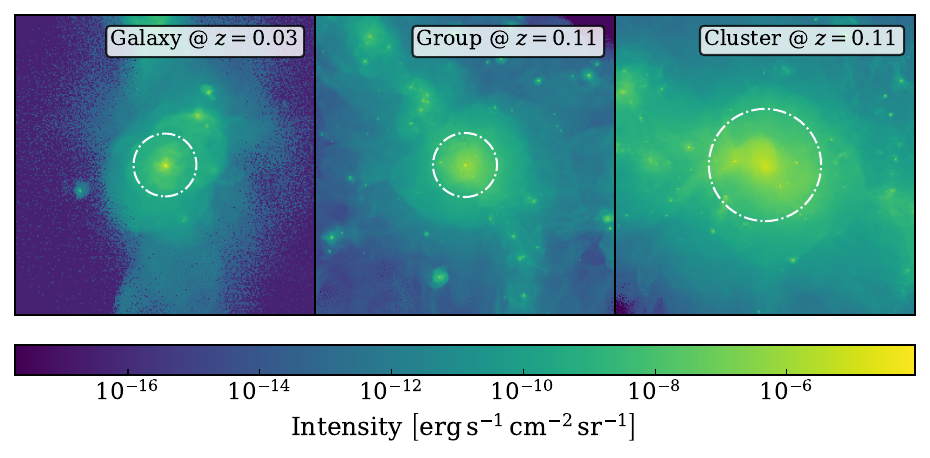}
  \caption{X-ray emissivity maps of observation targets only; i.e., generated using all gas particles bound to the target halos and no emissions from the foreground or background added to the map. The X-ray energy range is $\SIrange{0.1}{2}{keV}$. Panels from left to right: Galaxy with a total mass of $M = \SI{3.06e12}{M_{\sun}}$ at $z=0.03$, galaxy group with a total mass of $M = \SI{7.00e13}{M_{\sun}}$ at $z=0.11$, and galaxy cluster with a total mass of $M = \SI{3.78e14}{M_{\sun}}$ at $z=0.11$. Each panel is $256 \times 256$ in dimension, corresponding to a sky region of $1\degr \times 1\degr$. The white dash-dot circle denotes $R_{\mathrm{200c}}$ of the target.}
  \label{fig:targets}
\end{figure*}

In Fig.~\ref{fig:targets}, we present the X-ray emissivity maps of the selected targets. We used them as references to evaluate and compare emissions from light-cone intervening galaxies and diffuse gas along the lines of sight to these targets. In the figure, we label the $R_{\mathrm{200c}}$ (within which the mean density is equal to $200$ times the cosmological critical density) of the target sources with white dash-dotted circles. 

\begin{figure*}
  \centering
  \includegraphics[width=\linewidth]{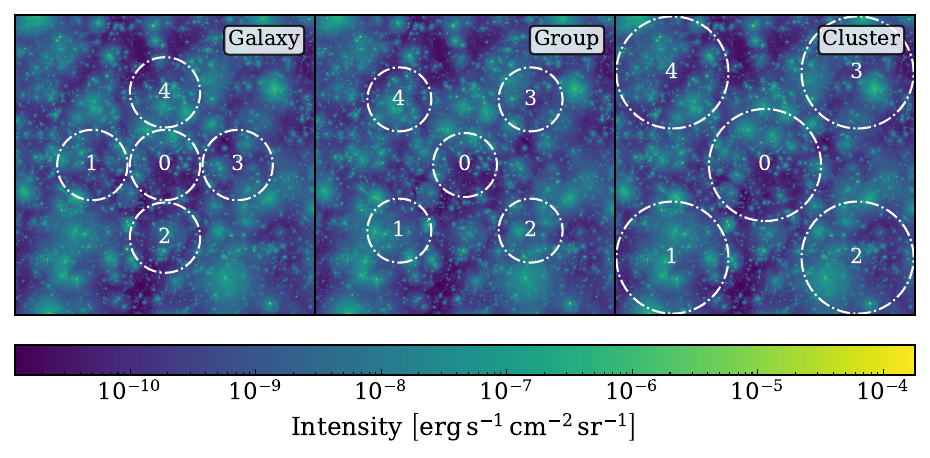}
  \caption{X-ray emissivity maps of the light cone with FoV of $1\degr \times 1\degr$. In each of the three panels from left to right, we selected five subregions (indicated by white dash-dotted circles) that have radii of $R_{\mathrm{200c}}$ corresponding to the target galaxy, group, and cluster (as given in Fig.~\ref{fig:targets}), respectively. }
  \label{fig:regions}
\end{figure*}

We refer to the emission of all intervening galaxies and diffuse gas in the light cone as the light-cone emission, which is calculated using the light cone (LC0) that we constructed up to $z=0.35$ from the simulation (see Sect.~\ref{sec:methods} for a detailed discussion). In order to study the average behavior and the variance of the light-cone emission, we selected five subregions, labeled R0, R1, R2, R3 and R4 in each case. Each subregion is set to have a radius of $R_{\mathrm{200c}}$ of the target source. The soft X-ray emissivity maps of the light cone are shown in Fig.~\ref{fig:regions}, where the three panels present the corresponding subregion locations in each of the target cases. 

\subsection{Hot-gas phase-space properties}

\begin{figure}
  \centering
  \includegraphics[width=\columnwidth]{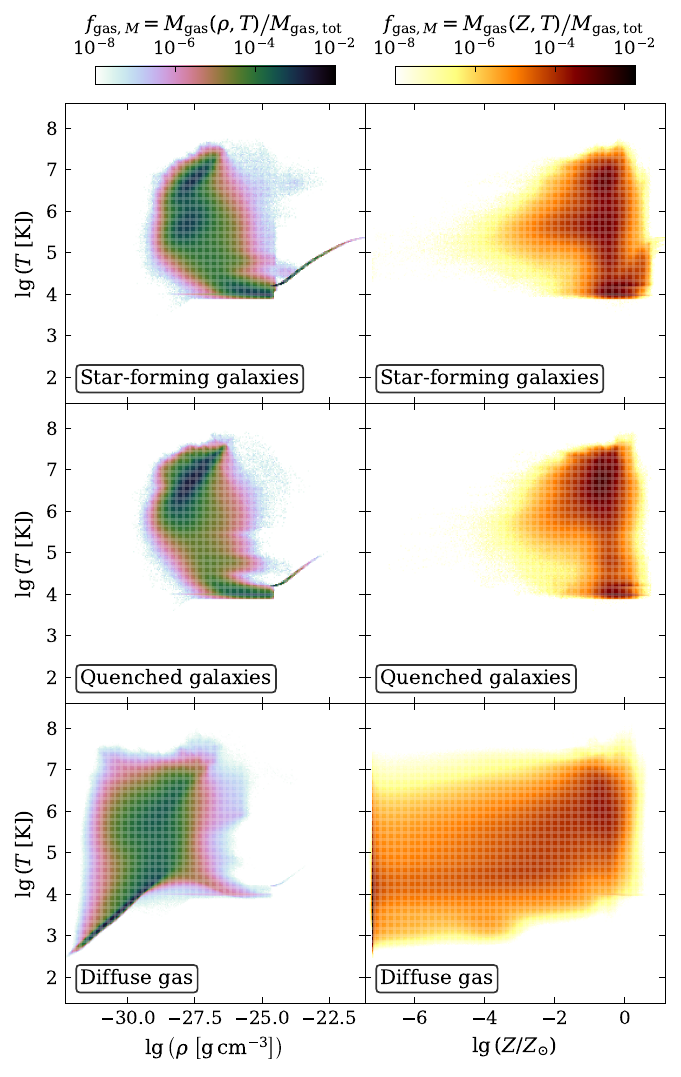}
  \caption{Phase-space diagrams of gas in the light cone. Left panels: Temperature versus density. Right panels: Temperature versus metallicity. Panels from top to bottom: Star-forming galaxies, quenched galaxies, and diffuse gas. The color maps represent the mass fraction of gas with respect to the total gas mass. }
  \label{fig:phase}
\end{figure}

Previous studies of the phase-space properties of hot gas in the simulation are limited to a single snapshot at the target source redshift \citep{2019MNRAS.484.5587T, 2019MNRAS.486.3766M, 2022MNRAS.510..399A}. In this work, we took an entire light cone into account, and in Fig.~\ref{fig:phase} we present the phase-space diagrams of the hot gas in different components of the light cone. These phase-space diagrams present 2D histograms of the mass fraction of gas with respect to the total gas mass in the density-temperature plane and in the metallicity-temperature plane, respectively. We made separate plots for the phase diagrams of gas residing in star-forming galaxies, quenched galaxies, and diffuse gas, respectively. 

As Fig.~\ref{fig:phase} shows, there are apparent differences between the phase diagrams of different kinds of gas. The long and narrow stripe feature in the diagram comes from star-forming gas, which is put on the effective equation of state by the multiphase star-forming subgrid model (\citealt{2003MNRAS.339..289S}). Star-forming galaxies contain more star-forming gas than the quenched galaxies, while the diffuse gas component contains the least star-forming gas, but it extends to much lower density regions. The temperatures of gas that is strongly gravitationally bound to galaxies are close to $\gtrsim \SI{e4}{K}$ with a clear density floor of $\sim \SI{e-29}{g.cm^{-3}}$ \citep{2012MNRAS.425.3024V}. The majority of gas in star-forming galaxies is in the temperature range of $\sim \SIrange[input-digits = 0123456789.]{e5}{e7.5}{K}$ and in quenched galaxies, $\sim \SIrange[input-digits = 0123456789.]{e6}{e7.5}{K}$. As galaxies are condense gas and metal reservoirs and severely suffer from chemical enrichment, the majority of gas in galaxies is located within a metallicity range of $\sim \SIrange{0.1}{1}{Z_{\sun}}$. The diffuse gas exhibits much lower metallicities. 

\subsection{Spectra}
\label{sec:spectra}

\begin{figure*}
  \centering
  \includegraphics[width=0.85\linewidth]{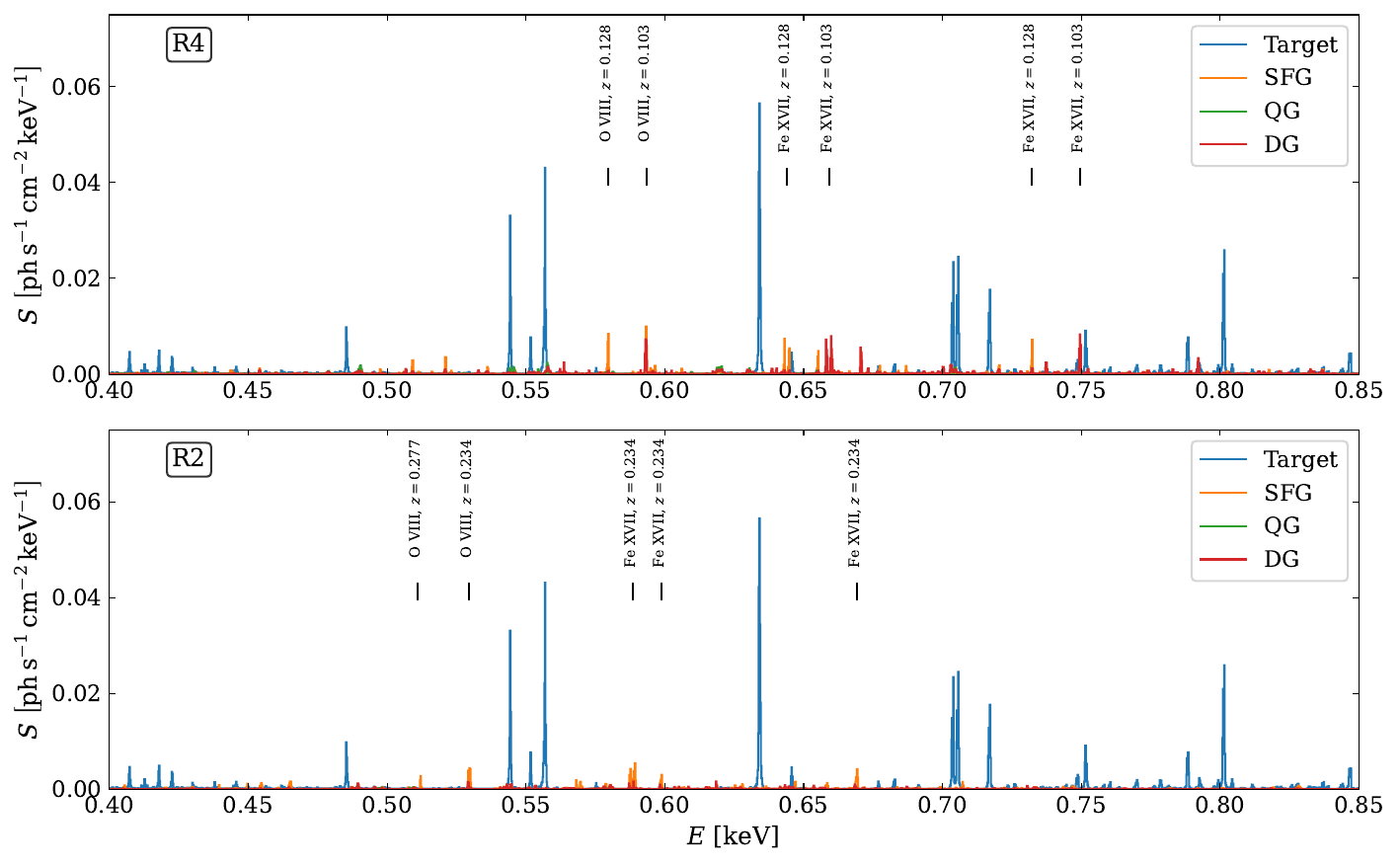}
  \caption{Two example combined spectra of target galaxy (blue) and light cone emission in two selected subregions (R2 and R4). The spectrum energy range shown here is $\SIrange{0.4}{0.85}{keV}$ and the bin width is $\SI{0.2}{eV}$. These were calculated within a radius of $R_{\mathrm{200c}}$ of the target galaxy and centered on the galaxy. The top panel presents spectra of a subregion (R4) with abundant strong emission lines from the light cone. The bottom panel shows spectra of a subregion (R2) with less abundant emission lines from the light cone. The labeled vertical black bars denote the most significant emission lines in the light cone, with identities and redshifts tagged. SFG represents star-forming galaxies, QG represents quenched galaxies, and DG represents diffuse gas. }
  \label{fig:spectra_galaxy}
\end{figure*}

We extracted signals from the selected regions to make the emission spectra. The example spectra (within the energy band that contains abundant strong emission lines) of the selected regions in the light cone are shown in Figs.~\ref{fig:spectra_galaxy}, \ref{fig:spectra_group}, and \ref{fig:spectra_cluster}. For comparison, the spectra of the targets were also plotted in these figures. In each case, we present two extreme situations in which the target is the most and least significantly contaminated by intervening emissions from the light cone. 

As can be seen, the spectra from the light cone contain abundant emission lines at different redshifts. In particular, strong emission lines in this given energy band include \ion{\ce{Fe}}{\romannumeral17} lines, \ion{\ce{O}}{\romannumeral7}, and \ion{\ce{O}}{\romannumeral8} lines. Observationally, when several (identified) emission lines correspond to the same redshift, they can be taken as coming from a common origin. In most cases, bright emission lines from the targets are much stronger than those from the intervening structures. In particular, in the case of galaxy clusters, because of the high temperature of the hot gas therein, continuum emission significantly dominates the spectra. However, there are cases where the X-ray emission from the intervening components along the line of sight becomes significantly strong (e.g., the top panels in Fig.~\ref{fig:spectra_group} and Fig.~\ref{fig:spectra_cluster}). We note that the soft X-ray spectra of star-forming galaxies contain more abundant emission lines than quenched galaxies (regarding this, we present a detailed quantitative comparison in Table~\ref{tab:landscape} from Sect.~\ref{sec:flux}). The spectra of diffuse gas contain the most abundant strong metal lines, especially \ion{\ce{Fe}}{\romannumeral17} lines. In addition, low-mass systems' galaxies tend to contribute more to the strong \ion{\ce{O}}{\romannumeral7} lines, while high-mass systems (e.g., massive groups and clusters) tend to generate stronger continuum and more abundant iron lines. We note that resolving emission lines properly is critical to distinguishing hot-gas emissions from the targets and from the intervening components along the lines of sight. 

\subsection{Fluxes within observed energy bands}
\label{sec:flux}

\begin{figure*}
  \centering
  \includegraphics[width=0.95\linewidth]{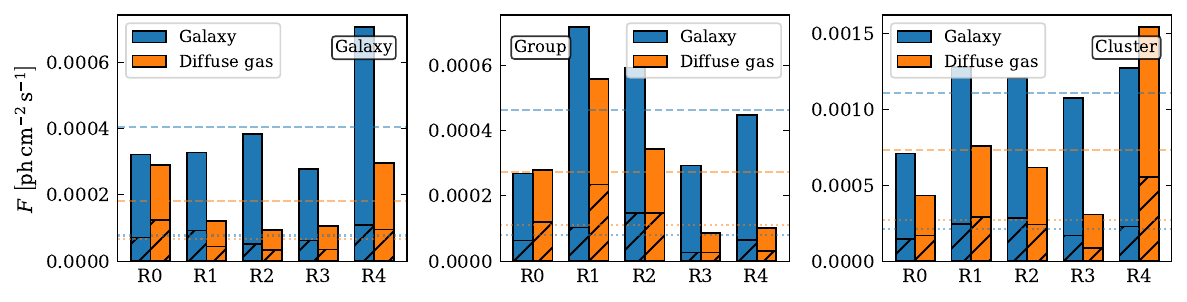}
  \caption{X-ray fluxes from line-of-sight galaxies (blue) and diffuse gas (orange) in different subregions. Panels from left to right correspond to the three cases: targeting a galaxy, a group, and a cluster (as given in Fig.~\ref{fig:targets}), respectively. The shaded areas represent the fluxes between $\SI{0.4}{keV}$ and $\SI{0.85}{keV}$. The dashed lines represent the mean values of the fluxes among five subregions in each case. The dotted lines represent the mean values in the shaded parts. }
  \label{fig:gal_diff}
\end{figure*}

In Table~\ref{tab:landscape}, we summarize the X-ray flux ratios between the light-cone emission (in the selected subregions) and the targets (both within $R_{\mathrm{200c}}$ of the target) within four different observed energy bands, including a full soft X-ray band in $\SIrange{0.1}{2}{keV}$, a narrower band in $\SIrange{0.4}{0.85}{keV}$ (as presented in Figs.~\ref{fig:spectra_galaxy}, \ref{fig:spectra_group} and \ref{fig:spectra_cluster}), and two narrow bands one around the strongest line (a resonance line with a centroid rest-frame energy of $\SI{0.574}{keV}$) in the \ion{\ce{O}}{\romannumeral7} triplet and one around \ion{\ce{O}}{\romannumeral8} of the target sources. For the emission lines, the line centroid was identified by \textsc{specutils} \citep{2022zndo...5911360E} and the flux was calculated using a line width of $\SI{4}{eV}$. As can be seen, despite the situation varying case by case, the overall light-cone emission in $\SIrange{0.1}{2}{keV}$ (as presented in the very last row in the table) can reach $\SIrange{10}{30}{\percent}$ of the target flux level, already indicating a non-negligible contribution from the line-of-sight hot gas to the total emission. The situation is more severe for the energy band of $\SIrange{0.4}{0.85}{keV}$ (as presented in the second-to-last row in the table). In the three cases that were used as the target sources in this study, the light-cone emission (as from intervening emitters alone) within this energy range can reach $\SIrange{30}{80}{\percent}$, $\SIrange{30}{200}{\percent}$, and $\SIrange{20}{40}{\percent}$ of the fluxes of the target galaxy, group, and cluster, respectively. In particular, the light-cone emission takes the lowest fraction in the case of cluster; this is simply because the target cluster has much stronger X-ray emission within this energy range. For narrow-band fluxes measured around the \ion{\ce{O}}{\romannumeral7} and \ion{\ce{O}}{\romannumeral8} lines of the target sources (presented in the third- and fourth-to-last rows of the table), the light-cone contamination can reach $\sim \SI{20}{\percent}$ in a few cases. However, in most cases, the fluxes in these oxygen lines from the targets are significantly higher than the intervening emission by at least one order of magnitude, suggesting that these oxygen line emissions (as measured within narrow bands) are better tracers of the hot-gas emission from the target sources, in comparison to that measured in wider energy bands. 

We note that the soft X-ray luminosities of dark-matter halos depend on the details of the baryonic physics adopted by the simulation (in particular AGN feedback) and thus would vary from halo to halo, and from simulation to simulation. As such, instead of providing relative flux ratios with respect to the selected target halos from the TNG100 simulation, we also present the absolute fluxes (in unit of $\si{ph.s^{-1}.cm^{-2}}$) of the light-cone hot-gas emissions in five selected subregions as well as within four different observed energy bands  in Tables~\ref{tab:galaxy}, \ref{tab:group}, and \ref{tab:cluster}, where the reader may also note the the mean and $1\sigma$ scatter, as well as the maximum and minimum values among the subsamples in order to obtain a statistical view of the level of light-cone emission.  

We further made a distinction between the X-ray emissions from star-forming galaxy halos, quenched galaxy halos, and diffuse gas in the light cone. We first present Fig.~\ref{fig:gal_diff}, which shows the X-ray fluxes of all line-of-sight galaxies versus intervening diffuse gas both projected within the selected subregions. From this figure, we can see that fluxes within the full band between $\SIrange{0.1}{2}{keV}$ from intervening galaxies are generally stronger than fluxes of diffuse gas in most cases. Whereas the fluxes within the energy band between $\SI{0.4}{keV}$ and $\SI{0.85}{keV}$ (as given under shaded areas), which contains strong oxygen and iron lines, are similar between the galaxy emission and diffuse gas emission. Figure~\ref{fig:sfg_qg} presents a comparison between line-of-sight star-forming and quenched galaxies within the subregions. As can be seen, star-forming galaxies contribute significantly more emissions than quenched galaxies in all cases. We note that cosmological simulations typically predict that star-forming galaxies are X-ray brighter than quiescent galaxies (at fixed stellar mass) (see \citealt{2020MNRAS.494..549T, 2020MNRAS.498.3061R, 2020ApJ...893L..24O}), as the former produce a significant amount of X-ray emission into the hot halo due to stellar feedback \citep[e.g.,][]{2016ApJ...818L..24S, 2020ApJ...903...32F}, while the latter are typically quenched via high-speed outflow driven by supermassive-black-hole feedback pushing hot gas out of the dark-matter halo. These simulation results are broadly consistent with high-resolution X-ray observations from \textit{Chandra} and \textit{XMM-Newton}, which detected luminous X-ray coronae around massive spiral galaxies and hotter halo gas around lower mass galaxies likely due to stellar feedback \citep[e.g.,][]{2011ApJ...737...22A, 2015ApJ...804...72B, 2017ApJ...848...61B, 2017ApJS..233...20L, 2018ApJ...855L..24L, 2019ApJ...882L..23D}. Stacked \textit{eROSITA} observations also revealed that the CGM around more massive or star-forming galaxies exhibit higher X-ray luminosities (out to $\SI{100}{kpc}$) than their lower mass or quiescent counterparts \citep{2022ApJ...936L..15C}. 

%%%%%%%%%%%%%%%%%%%%%%%%%%%%%%%%%%%%%%%%%%%%%%%%%%%%%%%%%%%%%%
\section{Discussion and conclusions}
\label{sec:DiscussionConclusions}

In this work, we used the TNG100 simulation to study the soft X-ray emissions from different intervening hot-gas components in a typical wide-field light cone and estimate their impacts on spectral and flux measurements toward X-ray-emitting galaxy, group, and cluster halos at lower redshifts. In particular, we split the intervening hot gas into three categories; that is, the hot gas that is gravitationally bound to either star-forming or quenched galaxies, and the diffuse gas which is more tenuously distributed permeating the cosmic web structures. The main conclusions of this paper are listed below. 
\begin{enumerate}
    
    \item The total mass fraction of the diffuse gas in the light cone is almost one order of magnitude higher than that of the gravitationally bound gas (as shown in Table~\ref{tab:gal}), which is broadly consistent with the result using the TNG300 simulation by \citet{2021A&A...649A.117G}. This diffuse component produces strong oxygen and iron line emissions at different redshifts in the soft X-ray band (as presented in Figs.~\ref{fig:spectra_galaxy}, \ref{fig:spectra_group}, and \ref{fig:spectra_cluster}). 

    \item Hot gas that is gravitationally bound to galaxy halos and that is diffusely distributed as the intergalactic medium residing in the cosmic web structures occupy different regions of the temperature-density-metallicity phase space. Diffuse gas can contribute nearly equally to the soft X-ray emission compared to the hot gas in star-forming galaxies along the line of sight, while quenched galaxy halos contribute the least to the hot-gas emission in the light cone (as shown in Table \ref{tab:landscape}, as well as in Fig.~\ref{fig:gal_diff} and Fig.~\ref{fig:sfg_qg}). 

    \item The soft X-ray spectra and fluxes of the intervening hot gas differ strongly case by case due to the large variance in the foreground and background hot-gas distributions in line with a target source (see Table~\ref{tab:landscape}). The fluxes from all of the line-of-sight emitters as measured in the energy band of $\SIrange{0.4}{0.85}{keV}$ can reach $\sim \SIrange{20}{200}{\percent}$ of the target emission, lower for clusters and higher for galaxies. The fluxes measured in narrow bands around the \ion{\ce{O}}{\romannumeral7} (r) and \ion{\ce{O}}{\romannumeral8} (K$\alpha$) are typically a few percent of the target emission, indicating that these line emissions (as measured within narrow bands) better present the hot-gas emission of the target sources, compared to that measured in wider energy bands. 

\end{enumerate}

This study targets at a theoretical investigation on the soft X-ray emission from line-of-sight origins. We note that the conclusions of this study are drawn using the TNG100 simulation. The quantitative details strongly depend on the implementation of hot-gas physics. In particular, the strong metal emissions of the diffuse gas in the cosmic web structures are results of galactic feedback processes, which can discharge metals far from galaxy centers. To better understand the impacts of feedback on the distribution and properties of the hot gas, we should further compare other numerical cosmological simulations (e.g., EAGLE and SIMBA which adopt different models. From the observation perspective, the pioneering X-ray telescopes in the planning stages, such as \textit{HUBS} and \textit{Super DIOS}, will provide us with valuable knowledge of the hot gas of different cosmic origins as well as deep insights into the baryonic cycle in galactic ecosystems. 

%%%%%%%%%%%%%%%%%%%%%%%%%%%%%%%%%%%%%%%%%%%%%%%%%%%%%%%%%%%%%%
\begin{acknowledgements}
The authors thank the anonymous referee for the helpful comments and suggestions. The authors sincerely thank Dr. Junjie Mao for detailed and useful comments. The authors acknowledge the Tsinghua Astrophysics High-Performance Computing platform at Tsinghua University for providing computational and data storage resources that have contributed to the research results reported within this paper. 
\end{acknowledgements}

\bibliographystyle{aa}
\bibliography{refs}

\begin{appendix}

\onecolumn

\section{Combined spectra of the targets and the light cone emission in selected subregions}

\begin{figure*}
  \centering
  \includegraphics[width=0.85\linewidth]{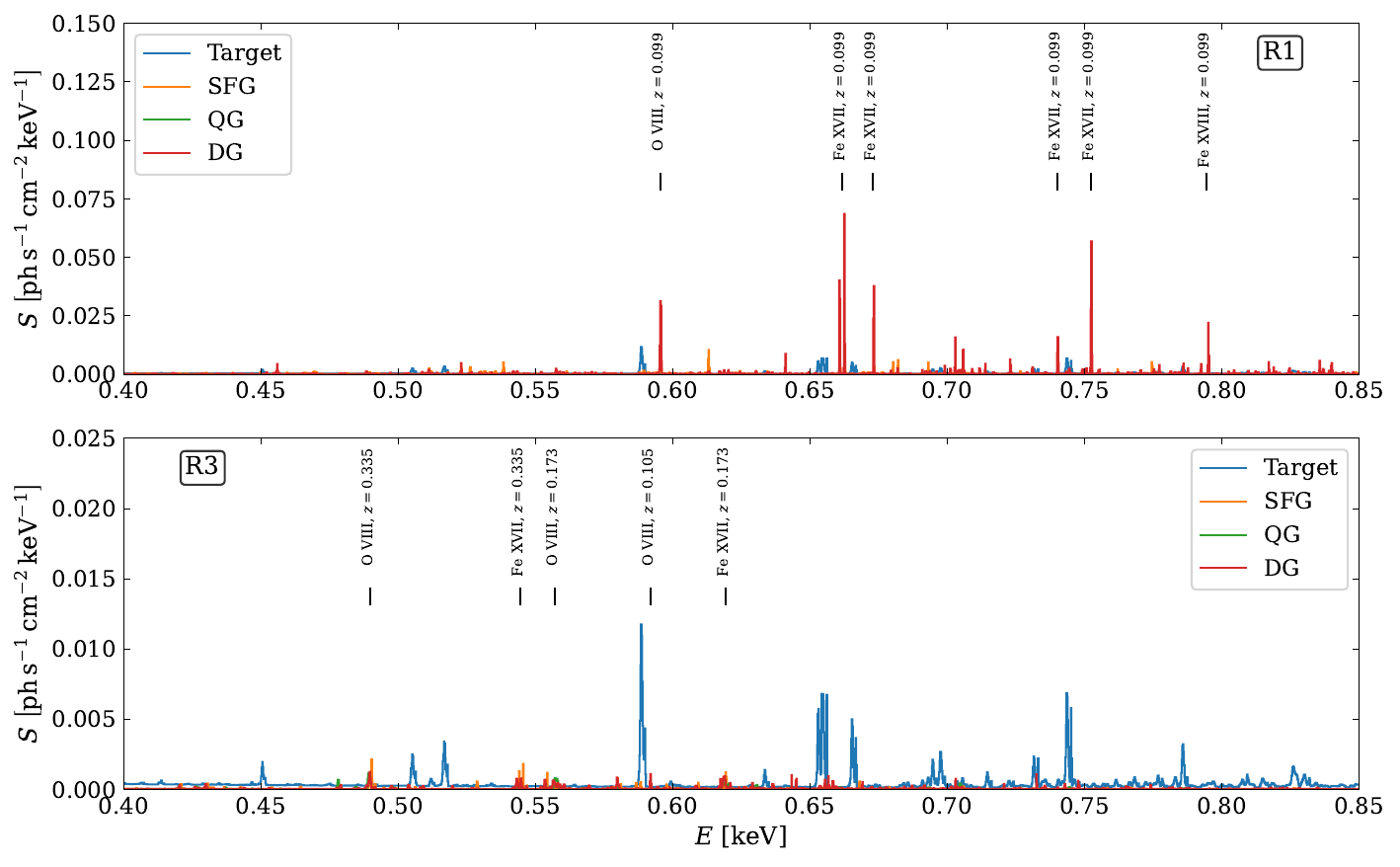}
  \caption{Same as Fig.~\ref{fig:spectra_galaxy}, but for the target group. we note that the emission from the target group (blue) is the same in both panels. Subregion R1 has much more significant light cone emission than subregion R3. }
  \label{fig:spectra_group}
\end{figure*}

\begin{figure*}
  \centering
  \includegraphics[width=0.85\linewidth]{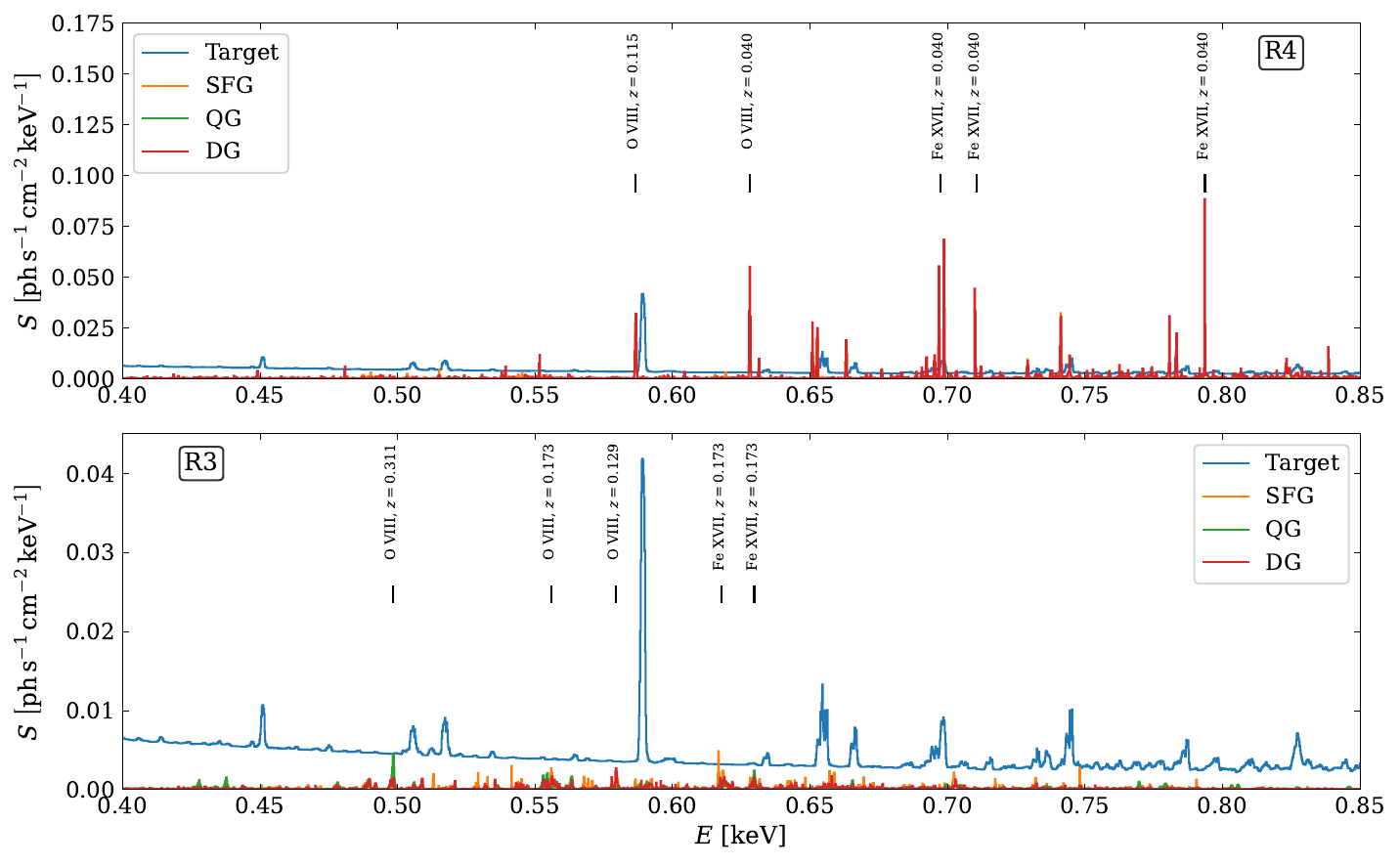}
  \caption{Same as Fig.~\ref{fig:spectra_galaxy}, but for the cluster. We note that the emission from the target cluster (blue) is the same in both panels. Subregion R4 has much more significant light cone emission than subregion R3. }
  \label{fig:spectra_cluster}
\end{figure*}

In Figs.~\ref{fig:spectra_group} and \ref{fig:spectra_cluster} we show the example spectra of the regions selected for the target group and the target cluster in the light cone, which are discussed in Sect.~\ref{sec:spectra}. 

\FloatBarrier

\section{X-ray fluxes in different subregions}

\begin{figure*}
  \centering
  \includegraphics[width=0.95\linewidth]{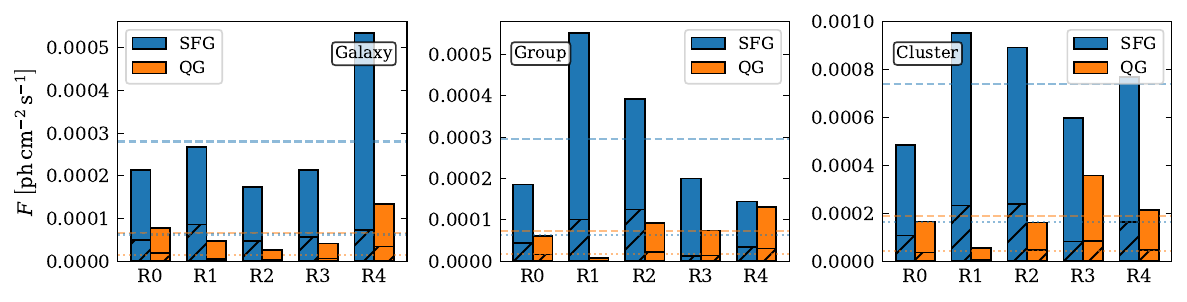}
  \caption{Same as Fig.~\ref{fig:gal_diff}, but for a comparison between line-of-sight star-forming galaxies (blue) and quenched galaxies (orange). }
  \label{fig:sfg_qg}
\end{figure*}

In Fig.~\ref{fig:sfg_qg} we show the comparison between line-of-sight star-forming and quenched galaxies within the subregions of the light cone.
As discussed in Sect.~\ref{sec:flux}, star-forming galaxies contribute significantly more soft X-ray emissions than quenched galaxies.

\FloatBarrier

\section{X-ray fluxes of the different subregions in the light cone}

In Table~\ref{tab:landscape} we report the X-ray flux ratios between the light-cone emission (in the selected subregions) and the targets (both within $R_{\mathrm{200c}}$ of the target) within four different observed energy bands, including a full soft X-ray band in $\SIrange{0.1}{2}{keV}$, a narrower band in $\SIrange{0.4}{0.85}{keV}$, and two narrow bands one around the strongest line (a resonance line with a centroid rest-frame energy of $\SI{0.574}{keV}$) in the \ion{\ce{O}}{\romannumeral7} triplet and one around \ion{\ce{O}}{\romannumeral8} of the target sources. 

In Tables~\ref{tab:galaxy}, \ref{tab:group}, and \ref{tab:cluster}, we report the absolute fluxes (in unit of $\si{ph.s^{-1}.cm^{-2}}$) of the light-cone hot-gas emissions in five selected subregions as well as within the four different observed energy bands.

\begin{table*}
 \scriptsize
 \centering
 \caption{X-ray flux ratios between the selected regions in the light cone and the targets.}
 \label{tab:landscape}
 \begin{tabular}{cl|S[table-format = 2.2]S[table-format = 2.2]S[table-format = 2.2]S[table-format = 2.2]S[table-format = 2.2]|S[table-format = 3.2]S[table-format = 3.2]S[table-format = 3.2]S[table-format = 2.2]S[table-format = 2.2]|S[table-format = 2.2]S[table-format = 2.2]S[table-format = 2.2]S[table-format = 2.2]S[table-format = 2.2]}
   \hline
   \multicolumn{2}{c|}{\multirow{3}{*}{Type}} & \multicolumn{5}{c|}{Galaxy} & \multicolumn{5}{c|}{Group} & \multicolumn{5}{c}{Cluster} \\
   \cline{3-17}
    & & {R0} & {R1} & {R2} & {R3} & {R4} & {R0} & {R1} & {R2} & {R3} & {R4} & {R0} & {R1} & {R2} & {R3} & {R4}  \\
    & & \multicolumn{5}{c|}{flux ratio between the light cone and the target} & \multicolumn{5}{c|}{flux ratio between the light cone and the target} & \multicolumn{5}{c} {flux ratio between the light cone and the target} \\
   \hline
   \multirow{4}{*}{\makecell{Star-\\forming\\galaxies}} & \ion{\ce{O}}{\romannumeral7} (r) & 0.72 & 1.07 & 0.41 & 1.27 & 4.83 & 3.13 & 10.53 & 4.21 & 0.23 & 11.67 & 1.29 & 2.40 & 2.14 & 2.15 & 7.23 \\
    & \ion{\ce{O}}{\romannumeral8} (K$\alpha$) & 0.64 & 1.18 & 0.91 & 1.46 & 1.47 & 2.61 & 14.33 & 7.64 & 2.29 & 5.12 & 0.97 & 9.87 & 3.57 & 1.28 & 1.39 \\
    & $\SIrange{0.4}{0.85}{keV}$ & 20.25 & 35.59 & 19.73 & 23.50 & 30.47 & 24.45 & 55.80 & 68.89 & 6.40 & 18.47 & 5.71 & 12.43 & 12.69 & 4.41 & 8.78 \\
    & $\SIrange{0.1}{2}{keV}$ & 7.15 & 8.97 & 5.83 & 7.17 & 17.97 & 3.88 & 11.56 & 8.22 & 4.19 & 3.02 & 4.12 & 8.09 & 7.57 & 5.07 & 6.53 \\
   \hline
   \multirow{4}{*}{\makecell{Quenched\\galaxies}} & \ion{\ce{O}}{\romannumeral7} (r) & 0.29 & 0.14 & 0.08 & 0.16 & 7.90 & 0.77 & 0.03 & 2.16 & 0.72 & 2.35 & 0.50 & 0.36 & 0.57 & 1.34 & 0.62 \\
    & \ion{\ce{O}}{\romannumeral8} (K$\alpha$) & 0.34 & 0.07 & 0.03 & 0.05 & 0.74 & 0.38 & 0.01 & 2.81 & 0.38 & 5.22 & 0.22 & 0.17 & 0.90 & 2.12 & 1.38 \\
    & $\SIrange{0.4}{0.85}{keV}$ & 8.10 & 2.34 & 1.51 & 2.47 & 14.25 & 8.65 & 0.39 & 11.85 & 7.50 & 16.84 & 1.93 & 0.30 & 2.47 & 4.54 & 2.43 \\
    & $\SIrange{0.1}{2}{keV}$ & 2.61 & 1.57 & 0.86 & 1.38 & 4.48 & 1.27 & 0.15 & 1.91 & 1.55 & 2.75 & 1.40 & 0.47 & 1.36 & 3.04 & 1.80 \\ 
   \hline
   \multirow{4}{*}{\makecell{Diffuse\\gas}} & \ion{\ce{O}}{\romannumeral7} (r) & 1.06 & 2.09 & 1.23 & 1.15 & 8.89 & 4.64 & 10.53 & 4.85 & 1.26 & 4.29 & 1.47 & 4.06 & 1.83 & 1.83 & 8.00 \\
    & \ion{\ce{O}}{\romannumeral8} (K$\alpha$) & 2.03 & 1.92 & 0.74 & 1.03 & 1.04 & 3.82 & 6.86 & 45.00 & 1.07 & 4.88 & 1.31 & 7.11 & 8.20 & 1.75 & 3.74 \\
    & $\SIrange{0.4}{0.85}{keV}$ & 51.32 & 18.44 & 13.63 & 14.34 & 39.77 & 66.27 & 129.44 & 81.67 & 14.62 & 17.01 & 9.03 & 15.58 & 12.83 & 4.71 & 29.63 \\
    & $\SIrange{0.1}{2}{keV}$ & 9.74 & 4.08 & 3.15 & 3.54 & 9.93 & 5.84 & 11.67 & 7.20 & 1.80 & 2.12 & 3.66 & 6.41 & 5.23 & 2.60 & 13.09 \\
   \hline
   \multirow{4}{*}{\makecell{All\\light cone\\emission}} & \ion{\ce{O}}{\romannumeral7} (r) & 2.11 & 3.38 & 1.73 & 2.57 & 21.66 & 8.62 & 21.27 & 11.23 & 2.25 & 18.31 & 3.27 & 7.12 & 4.56 & 5.44 & 16.08 \\
    & \ion{\ce{O}}{\romannumeral8} (K$\alpha$) & 3.04 & 3.18 & 1.68 & 2.54 & 3.28 & 6.84 & 21.22 & 55.45 & 3.74 & 15.23 & 2.51 & 17.54 & 12.71 & 5.15 & 6.54 \\
    & $\SIrange{0.4}{0.85}{keV}$ & 80.98 & 56.98 & 34.96 & 40.38 & 85.23 & 101.10 & 186.11 & 162.47 & 28.89 & 52.35 & 16.86 & 28.52 & 28.08 & 13.68 & 41.83 \\
    & $\SIrange{0.1}{2}{keV}$ & 20.58 & 15.10 & 16.06 & 12.88 & 33.73 & 11.44 & 26.69 & 19.60 & 7.93 & 11.47 & 9.67 & 17.26 & 15.46 & 11.72 & 23.87 \\
   \hline
 \end{tabular}
\end{table*}

%%%%%%%%%%%%%%%%%%%%%%%%%%%%%%%%%%%%%%%%%%%%%%%%%%%%%%%%%%%%%%

\begin{table*}
    \centering
    \caption{X-ray flux in unit of $\left(\si{ph.s^{-1}.cm^{-2}}\right)$ of the selected give subregions for the target galaxy in the light cone.}
    \label{tab:galaxy}
    \begin{tabular}{cl|S[table-format = 1.2e-1]S[table-format = 1.2e-1]S[table-format = 1.2e-1]S[table-format = 1.2e-1]S[table-format = 1.2e-1]S[table-format = 1.2(1.2)e-1, uncertainty-mode = separate]}
    \hline
    \multicolumn{2}{c|}{\multirow{1}{*}{Type}} & {R0} & {R1} & {R2} & {R3} & {R4} & {$\text{mean} \pm 1\sigma$} \\
    \hline
    \multirow{4}{*}{\makecell{Star-\\forming\\galaxies}} & \ion{\ce{O}}{\romannumeral7} (r) & 1.58e-7 & 2.35e-7 & 8.95e-8 & 2.77e-7 & 1.06e-6 & 3.63(3.53)e-7 \\
    & \ion{\ce{O}}{\romannumeral8} (K$\alpha$) & 2.33e-7 & 4.30e-7 & 3.30e-7 & 5.33e-7 & 5.34e-7 & 4.12(1.17)e-7 \\
    & $\SIrange{0.4}{0.85}{keV}$ & 4.87e-5 & 8.55e-5 & 4.74e-5 & 5.65e-5 & 7.32e-5 & 6.23(1.48)e-5 \\
    & $\SIrange{0.1}{2}{keV}$ & 2.13e-4 & 2.67e-4 & 1.74e-4 & 2.14e-4 & 5.25e-4 & 2.80(1.31)e-4 \\
    \hline
    \multirow{4}{*}{\makecell{Quenched\\galaxies}} & \ion{\ce{O}}{\romannumeral7} (r) & 6.40e-8 & 2.98e-8 & 1.86e-8 & 3.48e-8 & 1.73e-6 & 3.75(6.77)e-7 \\
    & \ion{\ce{O}}{\romannumeral8} (K$\alpha$) & 1.23e-7 & 2.54e-8 & 1.12e-8 & 1.74e-8 & 2.70e-7 & 8.95(9.93)e-8 \\
    & $\SIrange{0.4}{0.85}{keV}$ & 1.95e-5 & 5.62e-6 & 3.62e-6 & 5.94e-6 & 3.42e-5 & 1.35(1.17)e-5 \\
    & $\SIrange{0.1}{2}{keV}$ & 7.76e-5 & 4.68e-5 & 2.57e-5 & 4.10e-5 & 1.34e-5 & 6.49(3.82)e-5 \\
    \hline
    \multirow{4}{*}{\makecell{Diffuse\\gas}} & \ion{\ce{O}}{\romannumeral7} (r) & 2.32e-7 & 4.56e-7 & 2.69e-7 & 2.51e-7 & 1.94e-6 & 6.31(6.62)e-7 \\
    & \ion{\ce{O}}{\romannumeral8} (K$\alpha$) & 7.39e-7 & 6.98e-7 & 2.70e-7 & 3.75e-7 & 3.77e-7 & 4.92(1.90)e-7 \\
    & $\SIrange{0.4}{0.85}{keV}$ & 1.23e-4 & 4.43e-5 & 3.28e-5 & 3.45e-5 & 9.56e-5 & 6.61(3.67)e-5 \\
    & $\SIrange{0.1}{2}{keV}$ & 2.90e-4 & 1.21e-4 & 9.37e-5 & 1.05e-4 & 2.96e-4 & 1.81(0.92)e-4 \\
    \hline
    \multirow{4}{*}{\makecell{All\\light cone\\emission}} & \ion{\ce{O}}{\romannumeral7} (r) & 4.60e-7 & 7.38e-7 & 3.78e-7 & 5.63e-7 & 4.74e-6 & 1.38(1.69)e-6 \\
    & \ion{\ce{O}}{\romannumeral8} (K$\alpha$) & 1.10e-6 & 1.16e-6 & 6.11e-7 & 9.25e-7 & 1.19e-6 & 9.97(2.14)e-7 \\
    & $\SIrange{0.4}{0.85}{keV}$ & 1.95e-4 & 1.37e-4 & 8.40e-5 & 9.70e-5 & 2.05e-4 & 1.43(0.49)e-4 \\
    & $\SIrange{0.1}{2}{keV}$ & 6.13e-4 & 4.50e-4 & 4.78e-4 & 3.83e-4 & 1.00e-3 & 5.86(2.22)e-4 \\
    \hline
    \end{tabular}
\end{table*}

\begin{table*}
    \centering
    \caption{X-ray flux in unit of $\left(\si{ph.s^{-1}.cm^{-2}}\right)$ of the five selected subregions for the target group in the light cone.}
    \label{tab:group}
    \begin{tabular}{cl|S[table-format = 1.2e-1]S[table-format = 1.2e-1]S[table-format = 1.2e-1]S[table-format = 1.2e-1]S[table-format = 1.2e-1]S[table-format = 1.2(1.2)e-1, uncertainty-mode = separate]}
    \hline
    \multicolumn{2}{c|}{\multirow{1}{*}{Type}} & {R0} & {R1} & {R2} & {R3} & {R4} & {$\text{mean} \pm 1\sigma$} \\
    \hline
    \multirow{4}{*}{\makecell{Star-\\forming\\galaxies}} & \ion{\ce{O}}{\romannumeral7} (r) & 1.46e-7 & 4.92e-7 & 1.97e-7 & 1.07e-8 & 5.46e-7 & 2.78(2.06)e-7 \\
    & \ion{\ce{O}}{\romannumeral8} (K$\alpha$) & 3.58e-7 & 1.97e-6 & 1.05e-6 & 3.14e-7 & 3.04e-7 & 8.79(6.06)e-7 \\
    & $\SIrange{0.4}{0.85}{keV}$ & 4.42e-5 & 1.01e-4 & 1.25e-4 & 1.16e-5 & 3.34e-5 & 6.29(4.27)e-5 \\
    & $\SIrange{0.1}{2}{keV}$ & 1.85e-4 & 5.52e-4 & 3.93e-4 & 2.00e-4 & 1.44e-4 & 2.95(1.55)e-4 \\
    \hline
    \multirow{4}{*}{\makecell{Quenched\\galaxies}} & \ion{\ce{O}}{\romannumeral7} (r) & 3.61e-8 & 1.26e-9 & 1.01e-7 & 3.39e-8 & 1.10e-7 & 5.64(4.19)e-8 \\
    & \ion{\ce{O}}{\romannumeral8} (K$\alpha$) & 5.20e-8 & 1.80e-9 & 3.87e-7 & 5.17e-8 & 7.18e-7 & 2.42(2.75)e-7 \\
    & $\SIrange{0.4}{0.85}{keV}$ & 1.56e-5 & 6.97e-7 & 2.14e-5 & 1.36e-5 & 3.04e-5 & 1.64(0.98)e-5 \\
    & $\SIrange{0.1}{2}{keV}$ & 6.07e-5 & 6.99e-6 & 9.14e-5 & 7.40e-5 & 1.32e-4 & 7.29(4.07)e-5 \\
    \hline
    \multirow{4}{*}{\makecell{Diffuse\\gas}} & \ion{\ce{O}}{\romannumeral7} (r) & 2.17e-7 & 4.93e-7 & 2.27e-7 & 5.89e-8 & 2.00e-7 & 2.39(1.41)e-7 \\
    & \ion{\ce{O}}{\romannumeral8} (K$\alpha$) & 5.25e-7 & 9.42e-7 & 6.19e-6 & 1.48e-7 & 6.71e-7 & 1.69(2.26)e-6 \\
    & $\SIrange{0.4}{0.85}{keV}$ & 1.20e-4 & 2.34e-4 & 1.48e-4 & 2.64e-5 & 3.08e-5 & 1.12(0.78)e-4 \\
    & $\SIrange{0.1}{2}{keV}$ & 2.79e-4 & 5.57e-4 & 3.44e-4 & 8.58e-5 & 1.01e-4 & 2.73(1.73)e-4 \\
    \hline
    \multirow{4}{*}{\makecell{All\\light cone\\emission}} & \ion{\ce{O}}{\romannumeral7} (r) & 4.03e-7 & 9.95e-7 & 5.25e-7 & 1.05e-7 & 8.56e-7 & 5.77(3.19)e-7 \\
    & \ion{\ce{O}}{\romannumeral8} (K$\alpha$) & 9.40e-7 & 2.92e-6 & 7.62e-6 & 5.14e-7 & 2.09e-6 & 2.82(2.55)e-6 \\
    & $\SIrange{0.4}{0.85}{keV}$ & 1.83e-4 & 3.37e-4 & 2.94e-4 & 5.22e-5 & 9.47e-5 & 1.92(1.10)e-4 \\
    & $\SIrange{0.1}{2}{keV}$ & 5.46e-4 & 1.27e-3 & 9.36e-4 & 3.79e-4 & 5.48e-4 & 7.37(3.25)e-4 \\
    \hline
    \end{tabular}
\end{table*}

\begin{table*}
    \centering
    \caption{X-ray flux in unit of $\left(\si{ph.s^{-1}.cm^{-2}}\right)$ of the selected give subregions for the target cluster in the light cone.}
    \label{tab:cluster}
    \begin{tabular}{cl|S[table-format = 1.2e-1]S[table-format = 1.2e-1]S[table-format = 1.2e-1]S[table-format = 1.2e-1]S[table-format = 1.2e-1]S[table-format = 1.2(1.2)e-1, uncertainty-mode = separate]}
    \hline
    \multicolumn{2}{c|}{\multirow{1}{*}{Type}} & {R0} & {R1} & {R2} & {R3} & {R4} & {$\text{mean} \pm 1\sigma$} \\
    \hline
    \multirow{4}{*}{\makecell{Star-\\forming\\galaxies}} & \ion{\ce{O}}{\romannumeral7} (r) & 3.43e-7 & 6.39e-7 & 5.69e-7 & 5.74e-7 & 1.93e-6 & 8.10(5.67)e-7 \\
    & \ion{\ce{O}}{\romannumeral8} (K$\alpha$) & 7.81e-7 & 7.94e-6 & 2.87e-6 & 1.03e-6 & 1.12e-6 & 2.75(2.70)e-6 \\
    & $\SIrange{0.4}{0.85}{keV}$ & 1.07e-4 & 2.33e-4 & 2.37e-4 & 8.25e-5 & 1.64e-4 & 1.65(0.63)e-4 \\
    & $\SIrange{0.1}{2}{keV}$ & 4.86e-4 & 9.54e-4 & 8.93e-4 & 5.98e-4 & 7.71e-4 & 7.40(1.76)e-4 \\
    \hline
    \multirow{4}{*}{\makecell{Quenched\\galaxies}} & \ion{\ce{O}}{\romannumeral7} (r) & 1.32e-7 & 9.66e-8 & 1.53e-7 & 3.58e-7 & 1.65e-7 & 1.81(0.92)e-7 \\
    & \ion{\ce{O}}{\romannumeral8} (K$\alpha$) & 1.81e-7 & 1.40e-7 & 7.24e-7 & 1.71e-6 & 1.11e-6 & 7.72(5.89)e-7 \\
    & $\SIrange{0.4}{0.85}{keV}$ & 3.61e-5 & 5.52e-6 & 4.62e-5 & 8.49e-5 & 4.55e-5 & 4.37(2.54)e-5 \\
    & $\SIrange{0.1}{2}{keV}$ & 1.65e-4 & 5.52e-5 & 1.61e-4 & 3.58e-4 & 2.13e-4 & 1.90(0.99)e-4 \\
    \hline
    \multirow{4}{*}{\makecell{Diffuse\\gas}} & \ion{\ce{O}}{\romannumeral7} (r) & 3.91e-7 & 1.08e-6 & 4.89e-7 & 4.87e-7 & 2.13e-6 & 9.17(6.56)e-7 \\
    & \ion{\ce{O}}{\romannumeral8} (K$\alpha$) & 1.05e-6 & 5.72e-6 & 6.59e-6 & 1.40e-6 & 3.01e-6 & 3.56(2.24)e-6 \\
    & $\SIrange{0.4}{0.85}{keV}$ & 1.69e-4 & 2.91e-4 & 2.40e-4 & 8.81e-5 & 5.54e-4 & 2.69(1.58)e-4 \\
    & $\SIrange{0.1}{2}{keV}$ & 4.32e-4 & 7.56e-4 & 6.17e-4 & 3.07e-4 & 1.54e-3 & 7.31(4.34)e-4 \\
    \hline
    \multirow{4}{*}{\makecell{All\\light cone\\emission}} & \ion{\ce{O}}{\romannumeral7} (r) & 8.72e-7 & 1.90e-6 & 1.22e-6 & 1.45e-6 & 4.29e-6 & 1.94(1.22)e-6 \\
    & \ion{\ce{O}}{\romannumeral8} (K$\alpha$) & 2.02e-6 & 1.41e-5 & 1.02e-5 & 4.14e-6 & 5.26e-6 & 7.15(4.40)e-6 \\
    & $\SIrange{0.4}{0.85}{keV}$ & 3.15e-4 & 5.34e-4 & 5.25e-4 & 2.56e-4 & 7.83e-4 & 4.83(1.86)e-4 \\
    & $\SIrange{0.1}{2}{keV}$ & 1.14e-3 & 2.04e-3 & 1.82e-3 & 1.38e-3 & 2.82e-3 & 1.84(0.58)e-3 \\
    \hline
    \end{tabular}
\end{table*}

\end{appendix}

\end{document}